# Recognising Top-Down Causation

George Ellis, University of Cape Town

**Abstract**: One of the basic assumptions implicit in the way physics is usually done is that all causation flows in a bottom up fashion, from micro to macro scales. However this is wrong in many cases in biology, and in particular in the way the brain functions. Here I make the case that it is also wrong in the case of digital computers – the paradigm of mechanistic algorithmic causation - and in many cases in physics, ranging from the origin of the arrow of time to the process of quantum state preparation. I consider some examples from classical physics; from quantum physics; and the case of digital computers, and then explain why it this possible without contradicting the causal powers of the underlying micro physics. Understanding the emergence of genuine complexity out of the underlying physics depends on recognising this kind of causation. It is a missing ingredient in present day theory; and taking it into account may help understand such mysteries as the measurement problem in quantum mechanics:

## 1: The Theme

A key assumption underlying most present day physical thought is the idea that causation is bottom up all the way: particle physics underlies nuclear physics, nuclear physics underlies atomic physics, atomic physics underlies chemistry, and so on. Thus all the higher level subjects are at least in principle reducible to particle physics, which is therefore the only fundamental science; as famously claimed by Dirac, chemistry is just an application of quantum physics [1].

However there are many topics that one cannot understand by assuming this one-way flow of causation. The flourishing subject of social neuroscience makes clear how social influences act down on individual brain structure [2]; studies in physiology demonstrate that downward causation is necessary in understanding the heart, where this form of causation can be represented as the influences of initial and boundary conditions on the solutions of the differential equations used to represent the lower level processes [3]; epigenetic studies demonstrate that biological development is crucially shaped by the environment [4].

What about physics? In this essay I will make the case that top-down causation is also prevalent in physics, even though this is not often recognised as such. This does not occur by violating physical laws; on the contrary, it occurs through the laws of physics, by setting constraints on lower level interactions. Thus my theme is that the *foundational assumption that all causation is bottom up is wrong, even in the case of physics* [5]. Some writers on this topic prefer to refer to "contextual effects" or "whole-part constraints". These are perfectly acceptable terms, but I will make the case that the stronger term "top-down causation" is appropriate in many cases.

**Causation:** The nature of causation is highly contested territory, and I will take a pragmatic view:

***Definition 1*: Causal Effect** *If making a change in a quantity X results in a reliable demonstrable change in a quantity Y in a given context, then X has a causal effect on Y.*

***Example***: I press the key labelled "A" on my computer keyboard; the letter "A" appears on my computer screen.

Now there are of course a myriad of causal influences on any specific event: a network of causation is always in action. What we usually do is to have some specific context in mind where we keep almost all parameters and variables fixed, allowing just one or two remaining ones to vary; if they reliably cause some change in the dependent variable in that context, we then label them as "the cause". For example in the case of a digital computer, we have

(*The laws of physics, a specific computer structure, software loaded, data*) ➔ (*output*) (1)

Now in a particular run on a specific computer, the laws of physics do not change and the high level software loaded (e.g. Microsoft Word) will be fixed, so the above reduces to

$$(\textit{data}) \xrightarrow[\textit{MS-Word}]{} (\textit{output}) \qquad (2)$$

If however we load new high level software (e.g. now we run Photoshop) we will end up with a different relation than (2):

$$(data) \rightarrow\rightarrow\rightarrow (output) \quad (3)$$
*Photoshop*

Hence both the data and the software are causes of the output, in the relevant context. The laws of physics are also a cause, but we usually don't bother to mention this.

**Existence**: Given this understanding of causation, it implies a view on ontology (existence) as follows: I assume that physical matter (comprised of electrons, protons, etc.) exists. Then the following criterion for existence makes sense:

***Definition 2*: Existence** *If Y is a physical entity made up of ordinary matter, and X is some kind of entity that has a demonstrable causal effect on Y as per Definition 1, then we must acknowledge that X also exists (even if it is not made up of such matter).*

This is clearly a sensible and testable criterion; in the example above, it leads to the conclusion that both the data and the relevant software exist. If we do not adopt this definition, we will have instances of uncaused changes in the world; I presume we wish to avoid that situation.

**2: Hierarchy of scales and causation**
The basic hierarchy of physical matter is indicated in Table 1. Because it indicates physical scales, it is also an indication of levels of bottom-up causation, hence one can call it a hierarchy of causation. The computer hierarchy and life sciences hierarchy are rather different: those hierarchies are based on causation rather than scale.

| | **Domain** | **Scale** | **Mass** | **Example** |
|---|---|---|---|---|
| | | | | |
| **L17** | **Cosmology** | **$10^{26}$m** | **$10^{53}$kg** | **Observable Universe** |
| **L16** | **Large Scale Structures** | **$10^{23}$m** | **$10^{47}$kg** | **Great Attractor, Sloan Great wall** |
| **L15** | **Galaxy Clusters** | **$10^{22}$m** | **$10^{45}$kg** | **Virgo cluster, Coma cluster** |
| **L14** | **Galaxies** | **$10^{21}$m** | **$10^{42}$kg** | **M31, NGC 1300, M87** |
| **L13** | **Star clusters** | **$10^{20}$m** | **$10^{35}$kg** | **Messier 92, Messier 69** |
| **L12** | **Stellar systems** | **$10^{12}$m** | **$10^{30}$kg** | **Binaries, Solar system** |
| **L11** | **Stars** | **$10^{10}$m** | **$10^{30}$kg** | **Sun, Proxima Centauri, Eta Carinae** |
| **L10** | **Planets** | **$10^{9}$m** | **$10^{24}$kg** | **Earth, Mars, Jupiter** |
| **L9** | **Continents** | **$10^{7}$m** | **$10^{17}$kg** | **Africa, Australia** |
| **L8** | **Land forms** | **$10^{4}$m** | **$10^{8}$kg** | **Atlas mountains, Andes** |
| **L7** | **Macro objects** | **1m** | **10 kg** | **Rocks, chairs, computers** |
| **L6** | **Materials** | **$10^{-2}$m** | **$10^{-1}$kg** | **Conductors, Insulators, semi-conductors** |
| **L5** | **Molecules/ chemistry** | **$10^{-9}$m** | **$10^{-25}$kg** | **$H_20$, $SiO_2$, $C_6H_{12}O_6$, $C_9H_{13}N_5O_{12}P_3$** |
| **L4** | **Atomic physics** | **$10^{-10}$m** | **$10^{-26}$ kg** | **Hydrogen atom, Carbon atom** |
| **L3** | **Nuclear physics** | **$10^{-14}$m** | **$10^{-27}$ kg** | **Neutron, Proton, Carbon nucleus** |
| **L2** | **Particle physics** | **$10^{-15}$m** | **$10^{-33}$ kg** | **Quarks, electrons, gluons** |
| **L1** | **Quantum gravity** | **$10^{-35}$m** | | **Superstrings** |

**Table 1**: *The hierarchy of physical matter.*[1]

Now a core aspect of this hierarchy of emergent properties is that one needs different vocabularies and language at the different levels of description. The concepts that are useful at one level are simply inapplicable at other levels. The effective equations at the various levels are valid at a specific level but are valid for restricted conditions. They are written in terms of the relevant variables at those levels. These emergent variables can sometimes be obtained by coarse graining of lower level states, but not always so.

The key feature is that the higher level dynamics is effectively decoupled from lower level laws and details of the lower level variables [1]: with a few exceptions related to structured systems, you don't have to know those details in order to predict higher level behaviour.

[1] A fuller description is given at http://www.mth.uct.ac.za/~ellis/cos0.html.

A sensible view is that the entities at each classical level of the hierarchy (Table 1) are real. A chair is a real chair even though it is made of atoms, which in turn are real atoms even though they are made of a nucleus and electrons, and so on; and you too are real (else you could not read this paper), as is the computer on which you are reading it. Issues of ontology may be unclear at the quantum level, but they are clear at the macro level.

**3: Contextual effects in classical physics, astronomy, and cosmology**

Contextual effects occur commonly in classical physics. An example is nucleosynthesis in the early universe: macro-level variables (average cosmic densities) determine the expansion rate of the cosmos, which determines the temperature-time relation T(t) that in turn determines the rates of nuclear reactions (micro-level processes) and hence the outcome of nucleosynthesis, leading to macro variables such as the overall mass fraction of Helium and Lithium in the early Universe. Occurrence of this top-down effect is why we can use element abundance observations to constrain cosmological parameters [6].

More generally, top-down causation happens wherever boundary conditions and initial conditions determine the results. Environmental variables (a macro scale concept) act down to determine the values of physical fields locally:

(*Laws of physics, matter/field description, boundary/initial conditions*) ➔ (*output*) (6)

Hence for specific matter/fields we find

(*boundary/initial conditions*) ➔➔➔ (*output*) (7)
*matter/field properties*

Examples are standing waves; hearing the shape of a drum; and the outcomes of the reaction diffusion equation. Physicists often try to minimise such boundary effects through the idealisation of an isolated system; but no real system is isolated either in space or in time. There is nothing new in all this: it's just that we don't usually talk about this as top-down effects. It may be helpful to do so [3].

**Astronomy:** In the context of astronomy/astrophysics, there is a growing literature on contextual effects such as suppression of star formation by powerful active galactic nuclei [7]. This is a top-down effect from galactic scale to stellar scale and thence to the scale of nuclear reactions. Such effects are often characterised as feedback effects.

**Cosmology:** In cosmology, there are two venerable applications of the idea of top-down effects: Mach's Principle, and the Arrow of Time [8] In each case it has been strongly suggested that boundary conditions on the Universe as a whole are the basic cause of crucial local effects (the origin of inertia, and the local direction of time that characterizes increasing entropy).

The former is not now much discussed, but the latter has strong recent support [9,10]. Some will claim this is "nothing but" the combined effect of all the particles and fields in the very early universe. Well yes and no: the effect is indeed due to all those particles, but it depends crucially on the *specific relationship* between them. Change that relationship from smooth to very clumpy: exactly the same particles will be there, but will have a very different effect. It's the spatial distribution that matters: the relationship between them is the specific cause of the local physical effect of existence of a unique arrow of time. You cannot describe that higher level relationship in terms of the variables and scales relevant at the lower levels in Table 1. And the outcome is crucially important: life would not be possible without a well-established local arrow of time.

**4: Quantum Physics**

Top down effects occur in the context of quantum physics too [5]. Here are some examples:

**Band Structure**: The periodic crystal structure in a metal leads (via Bloch's theorem) to lattice waves, and an electronic band structure depending on the particular solid involved, resulting in all the associated phenomena resulting from the band structure [11]. The entire machinery for describing the lattice periodicity refers to a scale much larger than that of the electron, and hence is not describable in terms appropriate to that scale. Thus these effects all exist because of the macro level properties of the solid - the crystal structure - and hence represent top-down causation from that structure to the electron states. This can lead to existence of quasiparticles such as phonons that result from vibrations of the lattice structure; it also leads to Cooper pairs and hence to superconductivity. Because these are all based in top-down action, they are emergent phenomena in the sense that they simply would not exist if the macro-structure did not exist, and hence cannot be understood by a purely bottom-up analysis, as emphasized strongly by Laughlin [12].

**Caldeira-Leggett model:** The Caldeira-Leggett model is a model for a system plus heat reservoir, used for the description of dissipation phenomena in solid state physics [13]. Here the Lagrangian of the composite system **T** consisting of the system **S** of interest and a heat reservoir **B** takes the form

$$L_T = L_S + L_B + L_I + L_{CT} , \quad (8)$$

where $L_S$ is the Lagrangian for the system of interest, $L_B$ that for the reservoir (a set of non-interacting harmonic oscillators), and $L_I$ that for the interaction between them. The last term $L_{CT}$ is a `counter term', introduced to cancel an extra harmonic contribution that would come from the coupling to the environmental oscillators. This term represents a top-down effect from the environment to the system, because $L_I$ completely represents the lower-level interactions between the system and the environment. The effect of the heat bath is more than the sum of its parts when $L_{CT} \neq 0$, because the summed effect of the parts is given by $L_I$.

**State vector preparation**: State vector preparation is key to experimental set-ups, and is a non-unitary process because it can produce particles in a specific eigenstate from a stream of particles that are not in such a state. Indeed it acts rather like state vector reduction, being a transition that maps a mixed state to a pure state.

How can this non-unitary process happen in a way compatible with standard unitary quantum dynamics? The crucial feature is pointed out by Isham [14]: the outcome states are drawn from some collection $E_i$ of initial states by being selected by some suitable apparatus, being chosen to have some specific spin state in the Stern-Gerlach experiment; the other states are discarded. This happens in two basic ways: separation and selection (as in the Stern Gerlach experiment), which is unitary up to the moment of selection when it is not, or selective absorption (as in the case of wire polarisers), which continuously absorbs energy. This top-down effect from the apparatus to the particles causes an effective non-unitary dynamics at the lower levels, which therefore cannot be described by the Schrodinger or Dirac equations.

In such situations, selection takes place from a (statistical) ensemble of initial states according to some higher level selection criterion, which is a form of top-down causation from the apparatus to the particles. This is a generic way one can create order out of a disordered set of states, and so generate useful information by throwing away what is meaningless [5]. The apparatus is specifically designed to have this non-unitary effect on the lower level. This example is important in its own right, but also because it points the way to considering similar top-down effects in photon detectors [5]. This may be a crucial part of why actual measurements in a realistic context are non-unitary in character.

## 5: Complex structures: Digital Computers and the Brain

**Digital Computers** Structured systems such as a computer constrain lower level interactions, and thereby paradoxically create new possibilities of complex behaviour. For example, the specific connections between p-type and n-type transistors can be used to create NOR, NAND, and NOT gates [15]; these can then be used to build up adders, decoders, flip-flops, and so on. It is the specific connections between them that channels causation and so enable the lower level entities to embody such logic; the physical structure constrains the movement of electrons so as to form a structured interaction network. The key physical element is that the structure breaks symmetry (cf: [16]), thereby enabling far more complex behaviour than is possible in isotropic structures such as a plasma, where electrons can go equally in any direction.

However hardware is only causally effective because of the software which animates it: by itself hardware can do nothing. Both hardware and software are hierarchically structured [17], with the higher level logic driving the lower level events. The software hierarchy for digital computers is shown in Table 2.

| | |
|---|---|
| Level 7: | Applications programs |
| Level 6: | Problem oriented language level |
| Level 5: | Assembly language level |
| Level 4: | Operating system machine level |
| Level 3: | Instruction set architecture level |
| Level 2: | Microarchitecture level |
| Level 1: | Digital logic level |
| Level 0: | Device level |

**Table 2**: *The causal hierarchy in a digital computer system* [17].

Entering data by key strokes is a macro activity, altering macro variables. This acts down (effect T1) to set in motion a great number of electrons at the micro physics level, which (effect D1) travel through transistors arranged as logic gates at the materials level; finally (effect B1) these cause specific patterns of light on a computer screen at the macro level, readable as text. Thus we have a chain of top-down action T1 followed by lower level dynamical processes D1, followed by bottom up action B1, these actions composed together resulting in a same level effective macro action D2:

$$D2 = B1 \circ D1 \circ T1 \qquad (5)$$

This is how effective same-level dynamics D2 at the higher level emerges from the underlying lower level dynamics D1. This dynamics is particularly clear in the case of computer networking [18], where the sender and receiver are far apart. At the sender, causation flows downwards from the Application layer through the Transport, Network and Link layers to the Physical layer; that level transports binary code to the other computer through cable or wireless links; and then causation flows up the same set of layers at the receiver end, resulting in effective same-level transport of the desired message from source to destination. The same level effective action is not fictitious: it is a reliable determinable dynamic effect relating variables at the macro level of description at the sender and receiver. If this were not the case you would not be able to read this article, which you obtained by such a process over the internet. There was nothing fictitious about that action: it really happened. Emergent layers of causation are real [16,1,3].

The result is that the user sees a same-level interaction take place, and is unaware of all the lower levels that made this possible. This is information hiding, which is crucial to all modular hierarchical systems [19]. The specific flow of electrons through physical gates at the physical level is determined by whether the high level software is a music playing program, word processor, spreadsheet, or whatever: a classic case of top-down causation (the lower level interactions would be different if different software were loaded, cf. (2) and (3) above). Hence what in fact shapes the flow of electrons at the gate level is the logic of the algorithms implemented by the top level computer program [20,21].

Four crucial points emerge.

**A: Causal Efficacy of Non Physical entities:** Both the program and the data are non-physical entities, indeed so is all software. A program is not a physical thing you can point to, but by Definition 2 it certainly exists. You can point to a CD or flashdrive where it is stored, but that is not the thing in itself: it is a medium in which it is stored. The program itself is an abstract entity, shaped by abstract logic. Is the software "nothing but" its realisation through a specific set of stored electronic states in the computer memory banks? No it is not because it is the *precise pattern* in those states that matters: a higher level relation that is not apparent at the scale of the electrons themselves. It's a relational thing (and if you get the relations between the symbols wrong, so you have a syntax error, it will all come to a grinding halt). Furthermore it is not the same as any specific realisation of that pattern. A story can told (and so represented in sound mediated by air vibrations), printed in a book, displayed on a computer screen, attended to in ones mind, stored in one's memories; it is not the same as any of these particular representations, it in itself is an abstract thing that can be realised in any of these ways; and it is the same with computer programs: they are abstract entities that can be physically realised in many different ways, bit no particular one of them is the same as the story itself. They are its varied representations. This abstract nature of software is realised in the concept of *virtual machines*, which occur at every level in the computer hierarchy except the bottom one [17]. But this tower of virtual machines causes physical effects in the real world, for example when a computer controls a robot in an assembly line to create physical artefacts.

**B: Logical relations rule at the higher levels**: The dynamics at all levels is driven by the logic of the algorithms employed in the high level programs [20]. They decide what computations take place, and they have the power to change the world [21]. This abstract logic cannot be deduced from the laws of physics: they operate in a completely different realm. Furthermore the relevant higher level variables in those algorithms cannot be obtained by coarse graining any lower level physical states. They are not coarse-grained or emergent variables: they are assigned variables, with specific abstract properties that then mediate their behaviour.

**C: Underlying physics allows arbitrary programs and data:** Digital computers are universal computers. The underlying physics does not constrain the logic or type of computation possible, which Turing has shown is universal [22]. Physics does not constrain the data used, nor what can be computed (although it does constrain the speed at which this can be done). It enables the higher level actions rather than constraining them. The program logic dictates the course of things.

**D: Multiple realisability at lower levels.** The same high level logic can be implemented in many different ways: electronic (transistors), electrical (relays), hydraulic (valves), biological (interacting molecules) for example. The logic of the program can be realised by any of these underlying physical entities, which clearly demonstrates that it is not the lower level physics that is driving the causation. This *multiple realisability* is a key feature characterising top-down action [23]: when some high level logic is driving causation at lower levels, it does not matter how that logic is physically instantiated: it can be realised in many different ways. Thus the top-down map T1 in (5) is not unique: it can be realised both in different physical systems, and in different micro states of the same system.

**Equivalence classes:** This means that we can consider as the essential variables in the hierarchy, the *equivalence classes of lower level states* that all that correspond to the same high level state [23]. When you control a top level variable, it may be implemented by any of the lower level states that correspond to the chosen high level state; which one occurs is immaterial, the high level dynamics is the same. You even can replace the lower level elements by others, the higher entity remains the same (the cells in your body are now all different than they were 7 years ago; you are made up of different matter, but you are still the same person).

In digital computers, there are such equivalences all over the place:

- at the circuit level: one can use Boolean algebra to find equivalent circuits;
- at the implementation level: one can compile or interpret (giving completely different lower level functioning for same higher level outcome);
- at the hardware level, one can run the same high level software on different microprocessors;
- even more striking is the equivalence of hardware and software in much computing (there is a completely different nature of lower level entities for the same higher level outcomes).

In each case this indicates top-down effects are in action: the higher level function drives the lower level interactions, and does not care how it is implemented (information hiding is taking place).

Hence although they are the ultimate in algorithmic causation as characterized so precisely by Turing [22], digital computers embody and demonstrate the causal efficacy of non-physical entities. The physics allows this, it does not control what takes place. Computers exemplify the emergence of new kinds of causation out of the underlying physics, not implied by physics but rather by the logic of higher level possibilities. This leads to a different phenomenology at each of the levels of Table 2, described by effective laws for that level, and an appropriate language. A combination of bottom up causation and contextual affects (top-down influences) enables their complex functioning.

**Life and the brain**: living systems are highly structured modular hierarchical systems, and there are many similarities to the digital computer case, even though they are not digital computers. The lower level interactions are constrained by network connections, thereby creating possibilities of truly complex behaviour. Top-down causation is prevalent at all levels in the brain: for example it is crucial to vision [24,25] as well as the relation of the individual brain to society [2]. The hardware (the brain) can do nothing without the excitations that animate it: indeed this is the difference between life and death. The mind is not a physical entity, but it certainly is causally effective: proof is the existence of the computer on which you are reading this text. It could not exist if it had not been designed and manufactured according to someone's plans, thereby proving the causal efficacy of thoughts, which like computer programs and data are not physical entities.

## 6: Room at the bottom

Given this evidence for top-down causation, the physicist asks, how can there be room at the bottom for top-down causation to take place? Isn't there over-determination because the lower level physics interactions already determine what will happen from the initial conditions?

There are various ways that top down causation can be effective, based in lower level physical operations:

**a: Setting Constraints on lower level interactions**, which break symmetries [16] and so for example create more general possibilities than are available to cellular automata. This has been explained above in the context of digital computers. It is also the case where effective potentials occur due to local matter structuring [5]. These are crucially dependent on the nature of the higher level structure (e.g. crystal symmetries and gate structure) that cannot be described in terms of lower level variables. It is the higher level patterns that are the essential causal variable in solid state physics by creating a specific band structure in solids (hence for example the search for materials that will permit high temperature superconductivity). Which specific lower level entities create them is irrelevant: you can move around specific protons and electrons while leaving the band structure unchanged. This multiple realisability is always the case with effective potentials.

**b: Changing the nature of the constituent entities:** the billiard ball model of unchanging lower level entities underlying higher level structure is wrong. Hydrogen in a water molecule has completely different properties than when free; electrons bound in atom interact with radiation quite differently than when free; neutrons bound in nucleus have a half life of billions of years but they decay in 11 ½ minutes when free. Assuming that the nature of an entity is characterised by the way it interacts with others, in each case the higher level context has changed the nature of the underlying entities (an effect that is commonplace in biology).

**c: Creating constituent entities:** in many cases the lower level entities would not exist without the higher level structure. For example, the very possibility of existence of phonons is a result of the physical structure of specific materials [12]. This structure is at a higher level of description than that of electrons. Emergence of higher level entities has clearly occurred when lower level entities cannot exist outside their higher level context (again, a common effect in biology, where symbiosis is rife).

**d: Deleting lower level entities.** There is not a fixed set of lower level entities when selection creates order out of disorder by deleting unwanted lower level entities or states: top-down action selects what the lower elements will be. This selective top-down process is what underlies state vector preparation in quantum physics [5]. It is crucial in evolutionary biology [4] and in the mind [24,25].

**e: Statistical fluctuations and Quantum uncertainty:** Lower level physics is not determinate: random fluctuations and quantum indeterminism are both in evidence. What happens is not in the throes of iron determinism: random events take place at the micro level. In complex systems, this unpredictable variability can result in an ensemble of lower level states from which a preferred outcome is selected according to higher level selection criteria. Thus top-down selection leading to increased complexity [26] is enabled by the randomness of lower level processes.

## 7: The big picture: the nature of causation

The view put here is that the higher levels are causally real because they have demonstrable causal powers over the lower levels. Indeed the issue of ontology is clear at all higher levels: it is only at the lowest levels (where quantum effects dominate) that it gets murky.

I suggest top-down effects from these levels is the key to the rise of genuine complexity (such as computers and human beings):

**Hypothesis**: *bottom up emergence by itself is strictly limited in terms of the complexity it can give rise to. Emergence of genuine complexity is characterised by a reversal of information flow from bottom up to top down* [27].

The degree of complexity that can arise by bottom-up causation alone is strictly limited. Sand piles, the game of life, bird flocks, or any dynamics governed by a local rule [28] do not compare in complexity with a single cell or an animal body. The same is true in physics: spontaneously broken symmetry is powerful [16], but not as powerful as symmetry breaking that is guided top-down to create ordered structures (such as brains and computers). Some kind of coordination of effects is needed for such complexity to emerge.

The assumption that causation is bottom up only is wrong in biology, in computers, and even in many cases in physics, for example state vector preparation, where top-down constraints allow non-unitary behaviour at the lower levels. It may well play a key role in the quantum measurement problem (the dual of state vector preparation) [5]. One can bear in mind here that wherever equivalence classes of entities play a key role, such as in Crutchfield's computational mechanics [29], this is an indication that top-down causation is at play.

There are some great discussions of the nature of emergent phenomena in physics [17,1,12,30], but none of them specifically mention the issue of top down causation. This paper proposes that recognising this feature will make it easier to comprehend the physical effects underlying emergence of genuine complexity, and may lead to useful new developments, particularly to do with the foundational nature of quantum theory. It is a key missing element in current physics.

**Appendix**

The key idea is that of *functional equivalence classes*. Whenever you can identify existence of such equivalence classes, that is an indication that top-down causation is taking place [1]. Indeed this is essentially the ontological nature of the higher level effective entity: a computer program is in its nature the same as the set of all possible implementations of the set of logical operations it entails. These are what enter into the higher level effective relations; they can be described in many different ways, and implemented in many different ways; what remains the same in those variants is the core nature of the entity itself.

An *equivalence relation* is a binary relation ~ satisfying three properties:
For every element *a* in *X*, *a* ~ *a* (reflexivity),
For every two elements *a* and *b* in *X*, if *a* ~ *b*, then *b* ~ *a* (symmetry), and
For every three elements *a*, *b*, and *c* in *X*, if *a* ~ *b* and *b* ~ *c*, then *a* ~ *c* (transitivity).
The *equivalence class* of an element *a* is denoted [*a*] and may be defined as the set of elements that are related to *a* by ~.

The way this works is illustrated below: higher level states $H_1$ can be realised via variouss lower level states $L_1$. This may or may not result in coherent higher level action arising out of the lower level dynamics.

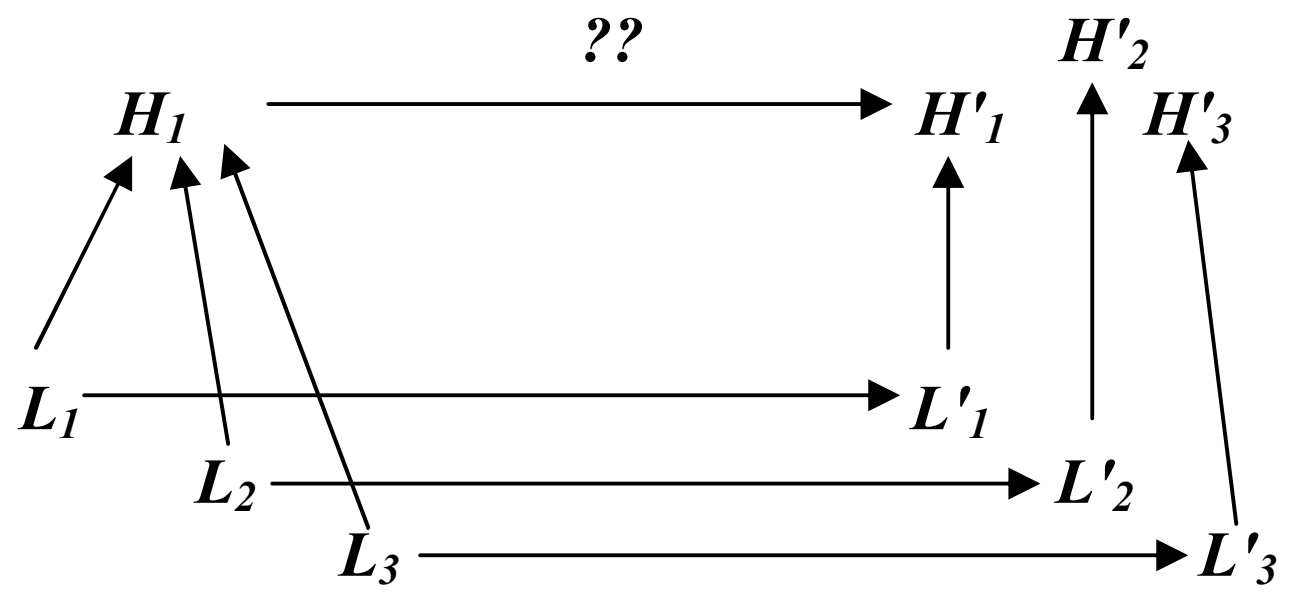

**Figure 1**: *The lower level dynamics does not lead to coherent higher level dynamics when the lower level dynamics acting on different lower level states corresponding to a single higher level state, give new lower level states corresponding to different higher level states.*

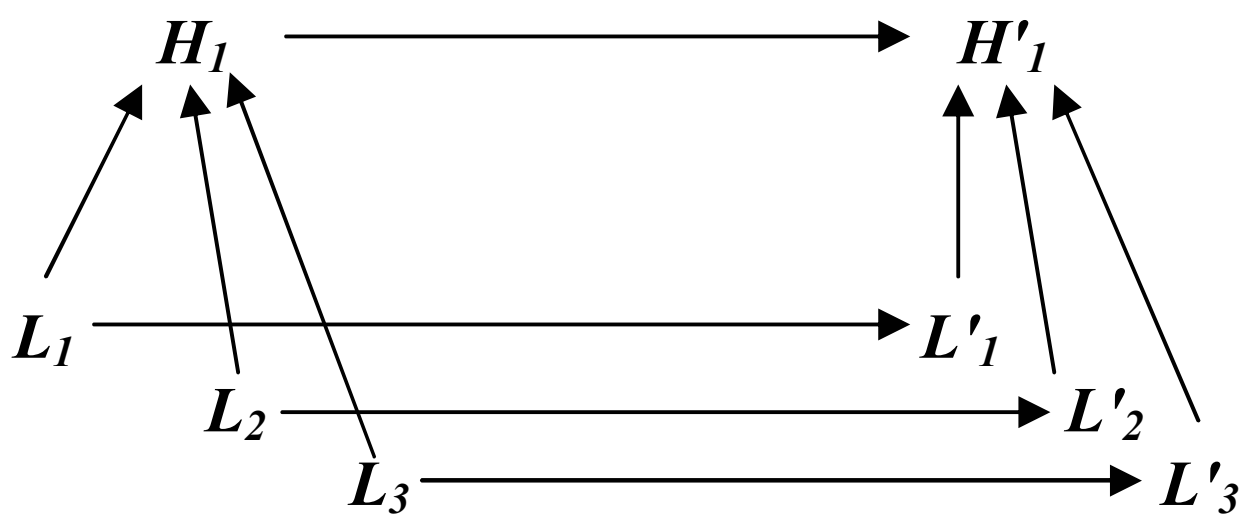

**Figure 2**: *The lower level dynamics leads to coherent higher level dynamics when the lower level dynamics acting on different lower level states corresponding to a single higher level state, give new lower level states corresponding to the same higher level state.*

When coherent dynamics emerges, the set of all lower states corresponding to a single higher level state form an equivalence class as far as the higher level actions are concerned. They are indeed the effective variables that matter, rather than the specific lower level state that instantiates the higher level one. This is why lower level equivalence classes are the key to understanding the dynamics.

**The ideal gas law** An example is the ideal gas law for gas constrained in a cylinder by a piston:

$$PV = n\,R\,T \qquad (4)$$

where the relation between the macroscopic variables pressure $P$, volume $V$, amount of gas $n$, and temperature $T$ does not depend on detailed microscopic variables such as velocities, positions and masses of any specific molecules; indeed we don't know those values. The gas is constrained by the cylinder and piston. The variables at the macro level are the only handle we have on lower level states: we can't (except in unusual circumstances) manipulate the micro level variables directly. But we can change them by altering macro variables (e.g. by compressing the gas, so changing $V$), which then changes lower level states (speeding up the molecules); this is the top down effect of the higher level variables on lower level states. The universal constant $R$ is the link between the micro and macro states, because it relates the energy of micro states to the values observed at the bulk level.

**Quantum cooperative effects** occur in superconductivity, superfluidity, and the quantum Hall effect. In superconductivity, the electrons - despite their repulsion for each other - form pairs ('Cooper pairs') which are the basic entities of the superconducting state. This happens by a cooperative process: the negatively charged electrons cause distortions of the lattice of positive ions in which they move, and the real attraction occurs between these distortions. The Nobel lecture by Laughlin [2] discusses the implications:

*"One of my favourite times in the academic year occurs in early spring when I give my class of extremely bright graduate students, who have mastered quantum mechanics but are otherwise unsuspecting and innocent, a take-home exam in which they are asked to deduce superfluidity from first principles. There is no doubt a very special place in hell being reserved for me at this very moment for this mean trick, for the task is impossible. Superfluidity, like the fractional Hall effect, is an emergent phenomenon - a low-energy collective effect of huge numbers of particles that cannot be deduced from the microscopic equations of motion in a rigorous way, and that disappears completely when the system is taken apart* [3] *... The students feel betrayed and hurt by this experience because they have been trained to think in reductionist terms and thus to believe that everything that is not amenable to such thinking is unimportant. But nature is much more heartless than I am, and those students who stay in physics long enough to seriously confront the experimental record eventually come to understand that the reductionist idea is wrong a great deal of the time, and perhaps always... The world is full of things for which one's understanding, i.e. one's ability to predict what will happen in an experiment, is degraded by taking the system apart, including most delightfully the standard model of elementary particles itself."*

The claim made here is that this dynamics is possible because of top-down causation.

Many other examples are to be found in [4]. It is implied in the work of Crutchfield [5], which is based in recognition of equivalence classes.